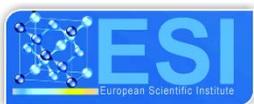

# Impact of Guided Inquiry with Simulations on Knowledge of Electricity and Wave Phenomena

**Fernando Espinoza Ed. D**

Department of Physics and Astronomy, Hofstra University, Hempstead, NY. USA.
Department of Chemistry and Physics-Adolescence Education, SUNY College at Old Westbury, Old Westbury, NY. USA

**Abstract**
The impact of inquiry-based instruction on the improvement of content knowledge of electricity and wave properties in two studies was the subject of this investigation. Groups of pre-college and undergraduate students performed a series of tasks designed to determine the effectiveness of exploratory/investigative approaches to the study of wave properties in fostering content retention and conceptual change, as well as in comparison with traditionally performed experimental activities. The results show that a guided inquiry-based method using simulations significantly improves content performance on electricity and wave motion, although not in optics; in particular, it helps students to visualize charge exchanges and electric neutrality, the shape of electric fields, types of waves and their characteristics, and the effects of the transmitting medium on the speed of a wave. Additionally, its exploratory nature is superior to the confirmatory laboratory experience, both in content retention and in facilitating the incorporation of perceptual features that help learners deal with documented challenging properties of waves.

**Subject:** Physics Education

**Introduction**

Content retention and its improvement are crucial measures of a learner's ability to benefit from further exposure to disciplinary core ideas, as declared by the Next Generation Science Standards (NGSS). An understanding of electricity, the properties of fields and wave phenomena play a large role in the study of physics and many areas of investigation in the sciences. In particular, the properties and relationships exhibited by waves can apply to phenomena that apparently have nothing in common, and can greatly enhance one's knowledge of the natural world by incorporating features such as dispersion, superposition, interference, and periodicity into investigations in other disciplines. The applications go beyond natural science; they exist in the social sciences as well, and in other areas of study such as economics in dealing with market forces and other human affairs.

Attempts to reform science education at a national level spanning several decades have led to recommendations from national organizations for instruction to emphasize exploratory, as opposed to confirmatory laboratory experiences (NAP, 2000). Subsequent analyses and recommendations argue that a lack of such exploratory opportunities seriously neglects students' pre-instructional knowledge and limits their engagement in meaningful experiences (Hofstein and Lunetta, 2004; America's Lab Report, 2006). There have been repeated calls for changes in teaching physics for more than a hundred years (Mann, C.R. 1906, 1912). While the vision expressed by these early critics remains largely unrealized (Otero and Meltzer, 2017), some progress has led to innovative instructional approaches. Indeed, a massive document-Adapting to a Changing World: Challenges and Opportunities in Undergraduate Physics Education (NAP, 2013) has identified specific needs and strategies for improving the learning and teaching of physics.

**Investigating Inquiry and its role in the study of Electricity and the Properties of Waves**

The most commonly used exploratory approach in science instruction is inquiry-based learning, as Figure 1 shows, there are degrees of such an instructional approach. There have been many studies on the effectiveness of inquiry-based learning, with some concluding that guidance is needed for successful implementation (Schneider, 2002; Kanter and Schreck, 2006; Krajcik et al. 2008; Minner et al. 2010). The effectiveness of inquiry-based learning on the development of critical thinking skills in several disciplines has been demonstrated at both the pre-college and undergraduate levels (DiPasquale et al. 2003; Abrahams and Millar, 2008; Casotti et al. 2008; Blanchard et al. 2010; Herrenkohl et al. 2011; Scalise et al. 2011; Redelman et al. 2012). The use of computer simulations to support inquiry-based instruction in learning science concepts has also been documented (Kirschner et al. 2006; Ryoo and Linn, 2012; Rutten et al. 2012; Chang and Linn, 2013; Pedaste et al. 2015). The advantages of teaching through inquiry have been questioned, and are debated in the literature (Quintana et al. 2005; Cobern et al. 2010; Rnekne and Nunez, 2013; Ronnebeck et al. 2016). However, a possible reason for much criticism may be a lack of awareness of the role of scaffolding made possible in many cases by computer simulations (Webb, 2005; Hmelo-Silver et al. 2007).

Teacher-Centered　　　　　　Guided Inquiry　　　　　　Student-Centered
　(Expository)　　　　　　　　　　　　　　　　　　　　(Exploratory)

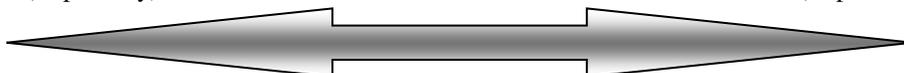

**Figure 1.** Representation of range of inquiry.

There are promising findings in using guided inquiry through interactive virtual settings in biology (Derting and Ebert-May, 2010), in physical science and chemistry (Donnelly et al. 2013; Chao et al. 2016), and in physics (Parnafes, 2007; Ceberio et al. 2016). The use of virtual environments (simulations) requires control, hence the need for scaffolding (Ding et al. 2011; Fang et al. 2016). Online simulations are challenging since quick and continuous changes place heavy demands on learners' working memory (particularly if prior knowledge is lacking); however, they also provide the user and the designer with two types of controls, pace control, and variable control. Pace control can facilitate comprehension, and variable control can engage the learner in knowledge construction, exploration, and hypotheses testing (Bjork and Linn, 2006). There exists a need for research on the use of open inquiry tasks to explore the effectiveness of variable manipulation (Wichmann and Timpe, 2015).

　　　　There is a documented lack of knowledge in the literature about high school students' ideas expressed in open-ended tasks, and very few empirical studies exist on the use of innovative online simulations such as the Electric Field Hockey simulation (Cao and Brizuela, 2016). Scientific reasoning can improve through the construction and analysis of models made possible by simulations, not just by their use (Heijnes et al. 2018). The Next Generation Science Standards emphasize modeling, and it is effective in increasing understanding of knowledge in certain domains (van Borkulo et al. 2011). Computers play a central role in studies using SimSketch, a modeling tool based on drawings in elementary school science (van Joolingen, 2012, 2015). The use of simulations in modeling is an integral part of developing inquiry processes, particularly student understanding of the relation between variables (White et al., 1999; van Joolingen et al., 2005). The use of simulations to support inquiry-based instruction has also been demonstrated to be effective by using Model Progression (Mulder et al., 2010, 2011, 2012). These studies have shown that novice learners can benefit by engaging in increasingly specific reasoning about the way variables are interrelated.

The increasing use of simulations and computer assisted learning in the teaching of physics has shown that there are compelling reasons for using authentic learning environments that result in more student engagement with the material (Murphy et al., 2006). Recent findings of the benefits of using inquiry-based instruction in the learning of thermal physics are particularly relevant since they come from Singapore, where policy makers are reconsidering traditional approaches in physics instruction (Fernandez, 2017).

Computer simulations are superior to traditional laboratory work in dealing with optical aberrations (Martinez et al. 2011), and the use of analogical scaffolding is more effective than traditional instruction in the study of electromagnetic waves (Podolefsky and Finkelstein, 2007).

Studies on student understanding of electromagnetism dealing with electrostatics and electric field properties have demonstrated a number of significant features (Viennot and Rainson, 1992; Tornkvist et al. 1993; Furio and Guisasola, 1998; Brian et al. 2009; Gire and Price, 2014); Bollen et al. 2017; Li and Singh, 2017; Klein et al. 2018). In particular, studies have focused on the use of the principle of superposition in electrostatics, exemplified by a force due to multiple charges (Bonham et al. 1999), and the role of computers on students' representations of electricity (Cronje and Fouche, 2008). Other studies have documented for several decades student conceptual difficulties with light propagation and geometric optics (Goldberg and McDermott, 1986, 1987; Perales Palacios et al. 1989; Bendall et al. 1993; Galili, 1996; Langley et al. 1997; Hubber, 2005; Bohm et al. 2009; Andreou and Raftopoulos, 2011; Lopes and Lopes, 2013).

Studies of student conceptualization of sound have documented that the mathematical abstraction (the syntax) seems divorced from the tangible character of its perceptual features (its semantic content). The latter based on a microscopic perspective that represents sound in terms of the bulk properties of a medium (Linder and Erickson, 1989). Student conceptualizations appear seemingly based on an awareness of disturbances through solids, with a predominance of air as the preferred medium. The benefits of conceptual change about sound propagation, using computer animations extend to students as young as 11 years of age (Calik et al. 2011).  In undergraduate physics instruction by virtue of its being primarily concerned with problem solving, such sense making (semantic content) is not readily apparent (Linder, 1993; Brown and Hammer, 2008). There is a need for research into student modeling of physical phenomena (Greca and Moreira, 2002), as well as the provision of scaffolding to facilitate student learning (Parnafes, 2010).

The particular focus of this study on electricity and mechanical waves as part of an inquiry-based approach to instruction is on the documented universality of students' use of concepts from kinematics and mechanics when they encounter the topics of electricity, and of waves. Another justification is the identification of some important cognitive features of student understanding of such phenomena. Studies in this area have gradually evolved from an identification of misconceptions to one of more comprehensive and structured patterns for student reasoning (diSessa and Sherin, 1998; Ambrose et al. 1999;McDermott and Redish, 1999; Wittmann et al. 1999).

The emphasis on inquiry-based instruction, and the epistemologically salient features of students' reasoning follow from studies where learners' reliance on explicit cognitive resources have been identified (Wittmann, 2002; Wittmann et al. 2003). These studies have effectively shown the inhibiting effect on learning caused by an adherence to object-like properties as opposed to extended or event-like ones. A proper understanding of field properties and wave characteristics, such as dispersion, superposition, and diffraction among others, tests the teleological explanations that often contain contradictory elements used by students, and that pose formidable challenges for instruction. This investigation specifically addresses two recommendations made in these studies:

1- The provision of learning environments alternative to traditional classroom tasks, so that student thinking seems more transparent through consistent reasoning is called for.

2- The need to explore alternative representations of electricity and wave properties.

Therefore, the research questions for the study became:
- To what extent do students succeed in demonstrating improvements in content knowledge of electrostatics, electric field and wave properties, when given the opportunity to engage in guided inquiry explorations using simulations?
- Are there sustainable benefits in content retention, when comparing control and experimental groups?
- What features of simulations lead to improvements in content performance?

To investigate the effectiveness of inquiry-based instruction on electrostatics, electric field and wave properties among both pre-college and undergraduate students, the study engaged several groups due to their similarity in conceptual level of development.

**Method**

The study based the similarity in conceptual level of development of the three groups (pre-college/Advanced Placement, non-science majors, and STEM majors) on the initial value of a pre-post measure of task performance using a normalized 'gain' determination. The measure of this gain reported in a previous study (Espinoza, 2015) with similar groups, resulted from tasks like that in Appendix A, as well as multiple-choice tests; the instruments were both the pre and post assessments of content performance. The normalized gain (h) formula $h = \frac{\% \, post - \% \, pre}{100 - \% \, pre}$ ensures that variability in pre-instructional knowledge levels is minimized among participants, by virtue of subtracting in the denominator the % pre-test from 100.

The measures used with the current groups reported elsewhere (Espinoza, 2016), were performances on pre-tests of content assessment. The pre-performance range for the groups was very similar to that of previous study groups, as well as subsequent measures of pre-performance on both tasks used in Study II with students in the same course as the one used as the control group.

The groups performed a series of guided-inquiry tasks designed to determine the effectiveness of exploratory/investigative approaches in fostering content improvement (study II), as well as a comparison of content retention with traditional confirmatory experimental activities (study I).

The rationale for conducting two studies was the need to engage sufficient numbers of students to have the means to investigate the two recommendations stated above. The undergraduate students taking the courses where they study wave properties do not study electricity and magnetism. Fortunately, in order to determine the effectiveness of simulations in the study of electricity, a different student population became available due to the collaboration with a physics teacher engaged in an independent study project with the author, for purposes of certification.

Study I

Two groups of undergraduates (the majority being second and third year students) were engaged in an investigation of the role of inquiry-based tasks in content retention of mechanical wave relationships expressed by the dependence of the speed on other variables. The study compared the use of a computer simulation for the first group, with the performance of a traditional laboratory task for the second group. The first group (the experimental one) was composed of non-science majors enrolled in a course on wave motion that fulfills a natural science requirement (N= 43).

The second group (the control one) was composed of STEM majors enrolled in a second semester introductory physics laboratory course (N= 34). The first group undertook an online exploratory activity (Appendix A) where they initially predicted the relationship between the frequency of oscillations created in a mechanical wave represented by a string or rope, and the transverse speed of the wave transmitted through it.

The activity specifically probed the documented tendency of students to attribute the speed of a mechanical wave to its source, and completely ignore the properties of the transmitting medium. Indeed, students deem the medium as passive, even useless, and leading them to conclude that a mechanical wave such as sound can travel through a vacuum (Maurines, 1992). The computer simulation designed to engage the students in confirming their prediction allowed for both pace and variable control. The questions included at the end of the activity summarize their findings about the relationship between wave speed in a string or rope and other variables, including wavelength, frequency, and tension.

The virtual exploratory task performance for this group was used as a comparison with a traditionally performed laboratory experiment on the laws governing vibrating strings, the latter being carried out by the STEM majors. The experiment calls for a confirmation of the relationship that exists between the fundamental frequency of a vibrating string and three variables: the string's length, its tension or stretching force, and its mass/unit length. The task incorporated two of the graphical relationships obtained in the experiment (frequency vs length, and frequency vs tension or stretching force), which are part of the laboratory report as questions into the pre and post assessment (Appendix B). The utility of the two questions used as the pre and post assessments rests on their widespread occurrence in textbooks and curriculum materials that deal with the study of mechanical waves. While there are only two questions, they are quite suitable and ideal for probing student understanding of the relationships between the variables that govern the behavior of vibrating strings. Other means of probing for an understanding of such a relationship may exist, but the use of the quantitative relationship expressed by an equation is economical and succinct.

Both groups received the same pre and post assessment. The pre assessments during the first meeting of the semester and the post assessments on the day of the final examination for both groups. The time intervals for both groups were two weeks between the pre assessment and the performance of the two tasks (the experiment with vibrating strings, and the assignment of the simulation); then there was an 11-week interval between the performance of the tasks and the final examinations. Care was exercised in planning for the groups to do both activities on the same week. The STEM majors performed the experiment in pairs, sharing the equipment and data collected; the non-science majors performed the simulation task on their own time during the two days between class meetings. There was no control over whether students received assistance, or if they worked alone in performing the task. However, given that the STEM majors worked in pairs, the sharing of information was a common occurrence. If anything, individual effort in performing the simulation adds to the benefit obtained from statistically significant differences in performance. Both groups dealt with the concepts involved in the tasks, although the non-science majors had less information, as their task contains a prediction. The STEM majors were predominantly second year students, whereas the non-science majors were predominantly third year students.

| Assessment | Activity | Completed | Control group (STEM majors) | Experimental group (non-science majors) |
|---|---|---|---|---|
| Pre-test | Content Knowledge | First meeting | √ | √ |
| Laboratory Report (two graphs) | Laboratory Experiment | Two weeks after pre-test | √ | |
| Project Report | Simulation | Two weeks after pre-test | | √ |
| Post-test | Content Knowledge | Final Exam 11 weeks after pre-test | √ | √ |

Table 1. Summary of activities and implementation

Table 1 provides a timeline and structure of the activities performed. The two questions included in the pre and post-test correspond to the tasks undertaken by both groups, but carried out differently. The control group (the STEM students) did them as part of their traditionally performed experiment. The experimental group (the non-science majors) did them as part of the exploratory task using a computer simulation designed to probe for the same relationships. In the assessment, question or task 1 tests for an understanding of the relationship between wave speed and other wave properties.

According to Espinoza (2017) wave speed **v** is expressed as $\mathbf{v} = \lambda f$, where $\lambda$ is wavelength, and $f$ is frequency. Additionally, task 2 tests for an understanding of the relationship between mechanical wave speed and properties of the transmitting material.

According to Serway (2019) a string's wave speed v is given by $\mathbf{v} = \sqrt{\dfrac{F}{\mu}}$, where **F** is the stretching force or tension in the string, and **μ** its mass/unit length. The objective is to establish a connection between the source of the mechanical waves, and the properties of the transmitting medium.

**Results**

| Semester | N | Pre | Sim. Task 1 | Sim. Task 2 | Gain (h) |
|---|---|---|---|---|---|
| **Fall 2016** | 22 | 43% | 76% | 76% | .58 |
| **Spring 2017** | 21 | 40% | 62% | 62% | .37 |

**Table 2**. Results of Wave task for the experimental group. The Pre measurements are the predictions, and the Sim. Tasks are the answers to the two questions after performing the simulation and having completed the table.

Table 2 shows the results of the exploratory activity for the experimental group with a gain in performance (*h*), using *% pre* = % of correct predictions at the beginning, and *% post* = % of correct answers at the end of the simulation task. In support of previously reported findings, the values of the gain (.58, and .37) fall in the medium-gain category of the graph of gain as a function of % pretest that compares traditional and interactive forms of engagement in the teaching of physics (Hake, 1998).

| Group | N | Task 1 | Post | Task 2 | Post |
|---|---|---|---|---|---|
| **Control** | 34 | 92% | 84% | 92% | 65% |
| **Experimental** | 43 | 69% | 68% | 69% | 59% |

**Table 3**. Performance for both groups on Tasks 1 and 2. The percentage for both tasks is the average grade for the laboratory report of the experiment for the control group, and the average grade for the report on the simulation task for the experimental group. Post is the answers to the same two questions used as the pre-assessment, given as part of the final examinations.

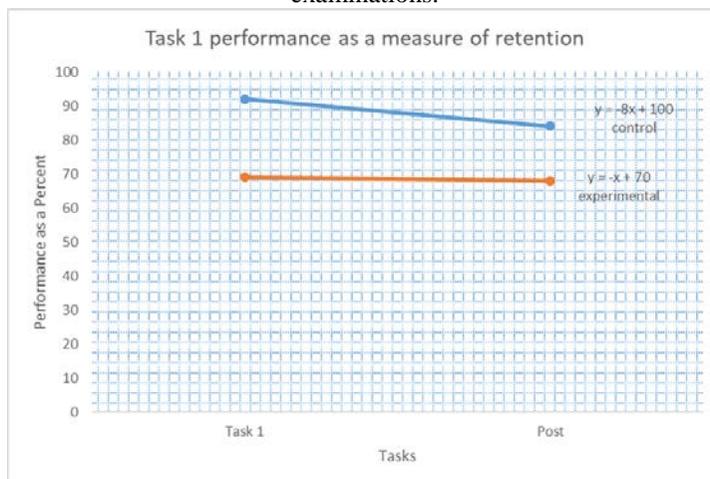

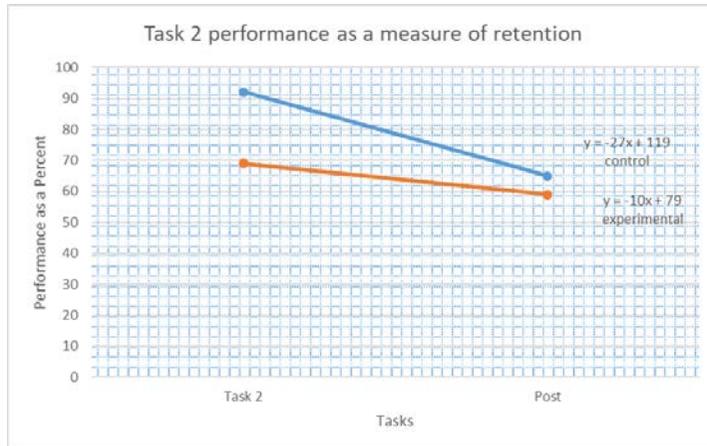

**Figure 2.** Graphs of performance on experimental tasks and post assessment for control and experimental groups.

Table 3 and the two graphs of content retention in Figure 2 measured in terms of a comparison of performance for both groups indicate that there is a difference between the scores obtained at two different times. For both groups there is a decrease in performance score from the time when they conducted the traditional experiment and the exploratory task, and the time of the post assessment. This corresponded to an approximately 11-week time span. Both groups undertook their respective tasks simultaneously and were administered the post assessment (as part of their final exam) 11 weeks later. It is clear that the experimental group outperformed the control group in the retention of information needed for the final exam, when they took the post assessment. The above conclusion is significant in light of the following results outlined in Table 4. The values for pre and post reported on the table are not means but the % of correct responses to the two questions.

| Group | N | Task 1 | | | χ2 | p value | Task 2 | | | χ2 | p value |
|---|---|---|---|---|---|---|---|---|---|---|---|
| | | Pre | Post | Gain (h) | | | Pre | Post | Gain (h) | | |
| **Control** | 34 | 68% | 84% | .50 | **.785** | **>. 10 (*)** | 54% | 65% | .24 | **7.45** | **< .01** |
| **Experimental** | 43 | 44% | 68% | .43 | | | 21% | 59% | .48 | | |

(*) Not statistically significant

Table 4. Comparison of performance by both groups on the same two questions (Pre: as the performance on the pre assessment before undertaking the tasks), and (Post: the answer to the same two questions as part of the final examinations). The values are not the means but the % of correct responses to the questions (Appendix B). The average gain for the control group was .37, and for the experimental group it was .46 for the two tasks.

Table 4 shows that there were significant gains for both groups between the pre and post assessment performances; in Task 1, they both fall in the medium-gain category of the graph of gain as a function of % pretest, in a meta-analysis that compares traditional and interactive forms of engagement in the teaching of physics (Hake, 1998). However, in Task 2, the gain is much higher for the experimental than for the control group, supporting the conclusions of that analysis that the more interactive exploratory task results in higher gain than the traditional experiment.

A chi-square analysis reveals no statistical significance between the groups' scores for Task 1, consistent with the first graph; however, the chi-square test $\chi2 = 7.45$, $p < .01$, yields a statistically significant difference between the two means for Task 2. The experimental group clearly outperforms the control group in the assessment of their grasp of the second relationship. The results suggest that while both groups benefit from exposure to experimental work in their understanding of the relationships between the properties of a mechanical wave, it is the exploratory or inquiry-based approach that seems superior in fostering student content retention, and a better grasp of perhaps the more challenging relationship between the variables (Task 2).

Study II

Variable groups of Advanced Placement (AP) physics students (N= 20-37) undertook tasks on electricity, electric field and wave properties using simulations to test their impact on performance on content knowledge of these topics. Although the research design for this study lacks the strength of the first one due to the lack of a comparison group, the opportunity provided by the collaboration between the author and the physics teacher was limited due to the term schedule and subject assignments of the teacher. The objective was to determine more precisely than in the first study, whether there is a statistically significant improvement in content performance due to engaging in guided inquiry simulations of the phenomena dealt with in the pre and post assessments. In addition, the post assessment probed for student feedback on their perceived impact of the simulations on the questions asked. The measurement tool for both pre and post assessments has a high validity score, as reported at the end of the article.

There were four topics addressed by the questions and the simulations: 1) electrostatics, 2) Electric Fields, 3) Wave Characteristics, and 4) Optics. The electrostatics pre/post assessment had 10 questions, with 37 students completing both the pre and posttest. Students completed an activity that explored the properties of static electricity using Phet simulations (Appendix C).

The electric fields pre/post assessment had 8 questions, with 24 students completing both the pre and posttest. Students completed an activity to explore the properties of electric fields and forces. Students used an online field line simulation to view and sketch field lines on point charges, and to play a game "Electric Field Hockey" to investigate the interaction of point charges within electric fields (Appendix D).

The wave characteristics pre/post assessment had 10 questions, with 21 students completing both the pre and posttest. Students completed an activity to explore the properties of waves using an online simulation to investigate properties such as frequency, amplitude, wavelength and speed, as well as the different motions of transverse and longitudinal waves. Students also used a Phet simulation to investigate sending a pulse down a string with a fixed end, and a loose end (Appendix E).

The optics pre/post assessment had 9 questions, with 20 students completing both the pre and posttest. Students investigated converging lenses using an online simulation (Appendix F).

In all four tasks, students completed the pretest, and then had about 30 minutes to complete the activity in class. Students completed the posttest in class the next day. All simulations used were HTML5 and accessible on students iPads. Some students who brought their laptops to school preferred to complete the simulations on their laptops, but overall, most students used their iPads. Very few students who did not have their iPads with them or had a broken iPad used their cell phones to complete the simulation.

**Results**

| Task | Performance | Pre | Post | Difference | t-test | P value | Cohen's d |
| --- | --- | --- | --- | --- | --- | --- | --- |
| **Electrostatics** | Mean | 37.88 | 49.39 | 11.52 | 25.71 | <.001 | .657 |
| | St. Dev. | 15.52 | 19.22 | 2.61 | | | |
| **Electric Fields** | Mean | 30.92 | 63.75 | 32.83 | 6.70 | <.001 | 1.98 |
| | St. Dev. | 12.51 | 20.05 | 26.19 | | | |
| **Wave Characteristics** | Mean | 43.33 | 64.76 | 21.43 | 5.28 | <.001 | 1.26 |
| | St. Dev. | 19.35 | 13.67 | 18.59 | | | |
| **Optics** | Mean | 30.56 | 28.72 | -1.83 | -.56 | .266 | -.147 |
| | St. Dev. | 13.98 | 10.46 | 13.84 | | | |

**Table 5.** Results of the performance on the four tasks; with the exception of the Optics task where the performance was quite regressive, the other tasks show a statistically significant

improvement in content performance, as well as large values of Cohen's d as a measure of Effect size.

| Question | Task | Yes (%) | No (%) |
|---|---|---|---|
| **Did the simulations Lab. Influence your answer choices For the post-test?** | Electrostatics | 31/33 (94) | 2/33 (6) |
| | Electric Fields | 20/20 (100) | 0/20 (0) |
| | Wave Characteristics | 18/21 (86) | 3/21 (14) |
| **What did you find most helpful About the simulations?** | Electrostatics:<br>• Visual exchanges of (+) and (-) charges<br>• Neutrality concept and attraction between charged and neutral objects<br>• Motion of electrons as compared to protons | | |
| **How did your answers change, if at all, after completing the simulations?** | Electric Fields:<br>• Direction of E<br>• (+) charges as sources, and (-) as sinks of E<br>Wave Characteristics:<br>• More clarity of Amplitude, wavelength, and frequency<br>• Understood vocabulary<br>• Longitudinal and transverse concepts | | |

**Table 6.** Results of student feedback on the post-test questions to determine their perspective on the role of the simulations in their performance. The post-test assessment for the Optics part did not include such questions.

Tables 5 and 6 provide a summary of the results of study II. As can be seen in Table 5, the differences in performance for the first three tasks are quite significant; as demonstrated here, for the electrostatics task a paired t-test of significance yields a value $t = 25.71$, $p < .001$, for the Electric Fields it is $t = 6.70$, $p < .001$, and for the wave characteristics $t = 5.28$, $p < .001$. The optics task was the exception, the performance being recessive. Table 6 contains the student feedback, and it provides information on the perceived reasons for the results. As the table shows, there is an overwhelming majority of students reporting that the simulations influenced their answer choices for the post-test. The remaining two questions provided specific features the students referred to as being helpful and decisive on their improvement in the post-test. This information is extremely useful for an analysis and reflection on improvements in other simulations in the same, and other content areas of physics instruction.

The assessment of proper questioning domain was done using the Lawshe test of content validity that yields a Content Validity Ratio (CVR) value (Lawshe, 1975). The findings show content validity ratios of the various sets of questions used as the pre and post assessments. Nine raters consisting of college and high school physics faculty from three institutions assessed the items used as measures of content performance.

| Task | Item Number | Lawshe CVR Ratio |
|---|---|---|
| **Electrostatics** | 10 | .76 |
| **Electric Fields** | 8 | .83 |
| **Wave Characteristics** | 10 | .84 |
| **Optics** | 9 | .90 |

**Table 7.** Results of the content validity analysis of the items used as the pre and post measures of performance.

Table 7 provides a summary of the results. The Electric Fields, Wave Characteristics, and Optics are at acceptable levels for the number of independent assessors of content validity (9 raters = minimum accepted scale value .78). The Electrostatics task barely misses the threshold value of content validity, perhaps due to the two items included that referred to charging by induction but containing diagrams, as opposed to another item dealing with induction but without a diagram. The remaining items dealt with charging by contact. Curiously, the Optics assessment received the highest CVR score from the raters, although no statistically significant performance for the students was found in this area.

The results confirm findings in studies that have found scaffolding to be effective in helping students understanding of electric field representations (Maries et al. 2017). The impact of the guided-inquiry simulations is also supportive of studies where guidance for discovery process provided significant gains in content knowledge of electric circuits (Kaakkola et al. 2011).

Additionally, the results are consistent with findings using pre and post measures that have found greater conceptual gains for innovative instruction, as opposed to traditional types (Dori and Belcher, 2005).

**Discussion**

The results from Study I suggest that perceptual features may be more evident for the experimental group in seeing how the changes in the tension of the string or rope make the difference in the wave speed. While both groups experience relatively similar gains in understanding the relationship expressed by $\mathbf{v} = \lambda\, f$, the results from the task involving $\mathbf{v} = \sqrt{\dfrac{\mathbf{F}}{\mu}}$ are much more significantly different between groups. It is precisely the second relationship that engages the learner in a deeper level of analysis when considering the effect of the transmitting medium on the properties of a mechanical wave. The initial differences in pre performance scores between the groups are readily understood in terms of prior knowledge differences between them. The control group is composed of students likely exposed in the lecture part of the course to the relationships before undertaking the tasks in the laboratory. Whereas the experimental group is encountering the questions as part of the course coverage, where there is no separation between lecture and laboratory parts. The findings in this study confirm and support those in other investigations; particularly those where students are found to initially think of mechanical waves in terms of the properties of objects rather than events, but eventually benefit in understanding and overcoming stubborn misconceptions by engaging in tasks where perceptual features of mechanical waves are made apparent (Wittmann et al. 2003).

The findings suggest reasons for the different outcomes in performance on Task 2 between the two groups, focusing on what the STEM majors (control) group did. The following graphs from the required laboratory report illustrate the likely reasons for the statistically significant difference in performance.

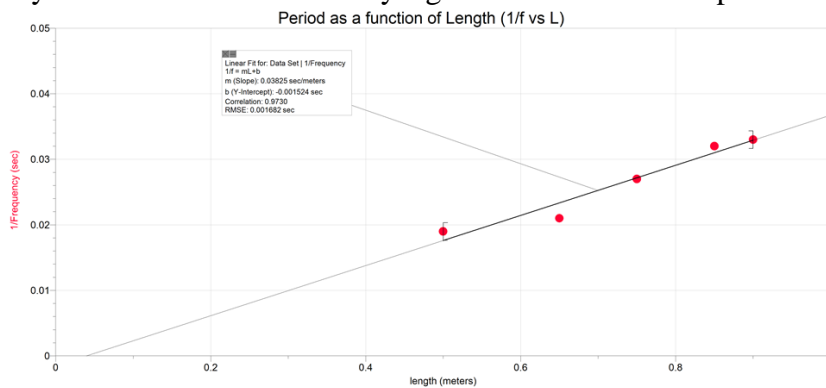

**Figure 3**. Task 1-Graph that translates into **v = λ *f*.** The first graphical relationship required as part of the laboratory report was obtained by a slight modification of the equation that incorporates the three independent variables (L, F, and μ) that affect the fundamental frequency of vibration of a string/wire, so that two experimentally obtained slopes can be compared to their theoretical values. The graph produced yields a relationship of the form y = mx, where only one variable is manipulated while keeping the other two constant. The relationship is as follows: (the values of m and μ are those used in the experiment)

$$f_1 = \frac{1}{2L}\sqrt{\frac{F}{\mu}} = \frac{1}{2L}\sqrt{\frac{mg}{\mu}} \tag{1}$$

Taking the reciprocal of the expression

$1/f_1 = 2L\sqrt{\dfrac{\mu}{mg}} = 2L\sqrt{\dfrac{.0011 \text{ kg/m}}{(.3\text{kg})(9.8 \text{ m}/sec^2)}}$, this equation is of the form y = mx

Where y = $1/f_1$, and m = $2\sqrt{\dfrac{.0011 \text{ kg/m}}{(.3\text{kg})(9.8 \text{ m}/sec^2)}}$ = .039 sec/m, is the theoretical slope

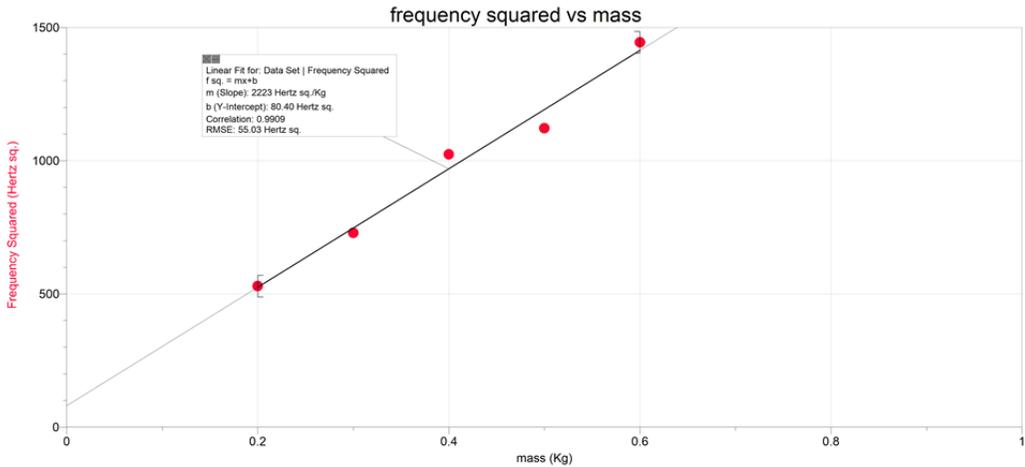

Figure 4. Task 2-Graph that translates into $\mathbf{v} = \sqrt{\dfrac{F}{\mu}}$. The second graph is as follows: (again using the values from the experiment)

Squaring the equation for the fundamental frequency $f_1$

$f_1{}^2 = \dfrac{mg}{4\mu L}2$, this equation is of the form y = mx, where y = $f_1{}^2$

and m = $\dfrac{g}{4\mu L}2 = \dfrac{9.8 \, m/sec^2}{4\,(1m)^2\left(\dfrac{.0011 Kg}{m}\right)} = 2{,}227$ Hz$^2$/ Kg is the theoretical slope.

Students likely find it easier to transition from the direct-linear graph ($1/f_1$ vs L) to the equation in the first question of the post assessment, than they do in the second one ($f_1{}^2$ vs m). Whereas the graph in Task 1 shows the same relationship between the frequency and the wavelength as does the question, this is not the case with the graph in Task 2 for the relationship between the frequency and the force/tension. The findings suggest that the students in the control group appear to handle much better an inverse than a power relationship between the variables.

　　　　As argued here, a caveat to the findings in the gains obtained by using the normalized (h) formula in Study I can be offered. There are concerns over the apparent bias that (h) contains in favor of high pre-test populations (Nissen et al., 2018). As demonstrated earlier, this was addressed in Study II by the inclusion of the evaluation of Cohen's d in addition to the t-test of significance.

Nevertheless, the significance of the greater improvement by the control group in Task 2 of the first study lies in the fact that they are the low pre-test group. Finally, the results of study II show a statistically significant impact of simulations on content performance improvement in electrostatics, electric field properties, and wave characteristics assessments. The Optics activity shows no improvement most likely due to a lack of correlation between the simulation task and the assessments (e.g., no mirror tasks in the simulation, whereas there were more than half such questions in the assessments). The robustness of the improvements receives support from the student feedback; it reduces the need to infer the likely reasons for improvement in content performance.

**Limitations and Recommendations**
While providing evidence of the positive impact on content performance due to inquiry-based activities where the tasks are student-centered, the studies did not address process skill development specifically; however, in terms of content retention it is apparent that embedding such practices in the tasks students engage in, leads to improvements in their content performance on the examined properties. It would be advisable to engage larger groups in tasks that can target both content, and process skill development in ways where one can more effectively extrapolate to general populations.

There could have been a potential influence on the performance by the non-science group due to the extra time available for them to undertake the simulation, which was not available to the STEM majors. This factor ought to receive consideration in a replicating study, to ensure a more rigorous methodological approach.

Researchers should test the effectiveness of the techniques by collecting data on students' understanding of other areas, such as mechanics. Nevertheless, the study is part of the awareness-raising call (Powell, 2003) to make available evidence that student-centered instruction helps in the development of scientific knowledge, as well as critical thinking skills.

**Conclusion**
The two studies provide evidence that students benefit from the use of simulations to improve their understanding of relationships between variables in situations where such relationships are not easily perceptible. There are statistically significant differences in improvement on content performance in electrostatics and electric field properties, as well as in several wave characteristics. The only area without improvement is likely due to the lack of correlation between the simulation tasks, and the assessment of content knowledge.

**Abbreviation**

STEM is an acronym for Science, Technology, Engineering, and Mathematics.

All procedures performed in studies involving human participants were in accordance with the ethical standards of the institutional and/or national research committee and with the 1964 Helsinki declaration and its later amendments or comparable ethical standards.

This article does not contain any studies with animals performed by the author. Informed consent was obtained from all individual participants included in the study.

The author wishes to thank Emily Ferrara for her assistance in the data collection, and Josh Fyman for the analysis of the content validity part.

**Appendix A**
**Exploratory Task**

Does the speed of a wave in a rope or a string depend on how fast the oscillations are created?
ANSWER:
What are your reasons for the answer?
Now test your answer by using a simulation at http://phet.colorado.edu/index.php
Choose "Wave on a string" from the available choices. With the simulation open, make sure it looks like the diagram below.

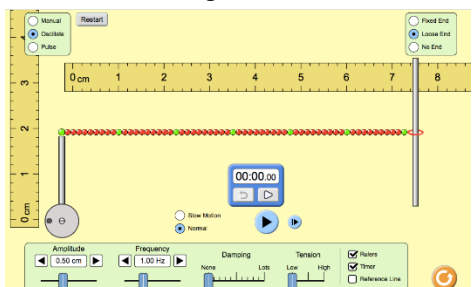

The horizontal ruler can be moved to determine the wavelength, while the vertical one stays put.
- Click *as simultaneously as possible* the Oscillate and Play/Start buttons and determine how long it takes for the wave to reach the ring at the end. Stop the timer and record the value. Determine the wavelength by placing the horizontal ruler so that it measures the distance between peaks. Repeat for "Frequency" (the equivalent of how fast the oscillations are set up) values of 2.0 and 3.0.

- Now select "Frequency" back to 1.0 and then change the tension to the middle of the scale; repeat the measurement of the time. Finally, change the tension to high and repeat once more, then fill in the table.

(*) The speed can be calculated by multiplying the wavelength by the frequency (v = λ f)

| Frequency (Hertz) | Tension | Timer Reading (seconds) | Wavelength (centimeters) | (*) Speed (cm/sec) |
|---|---|---|---|---|
| 1.0 | Low | | | |
| 2.0 | Low | | | |
| 3.0 | Low | | | |
| 1.0 | middle | | | |
| 1.0 | high | | | |

1- What does the table suggest for
a) The time taken for the waves to travel?
b) The speed of the waves?
2-a) What did you notice about the wavelengths as you made the changes?
b) Can you offer reasons for the answer in a)?

**Appendix B**

For the case of a string of length L (held fixed at both ends), a standing wave results when a transverse force is applied to the string.

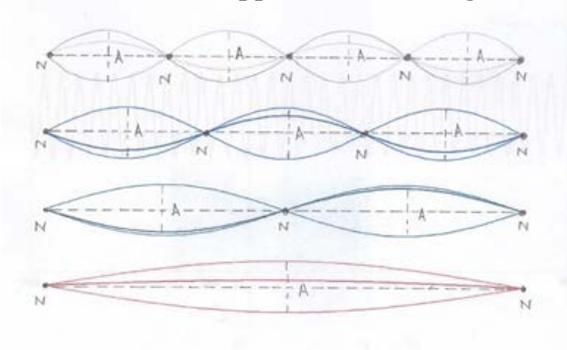

The bottom diagram represents the fundamental mode of vibration, and the remaining ones represent the harmonics or multiples of the fundamental. Since the speed of a wave is given by the product of its wavelength and frequency v = λ f. Since L = $\frac{\lambda}{2}$ (from the fundamental mode of vibration), Hence λ = 2L

Also, for a vibrating string $v = \sqrt{\dfrac{F}{\mu}}$ (the speed is the ratio of the tension or stretching force and the inertial mass/unit length property of the string). Therefore,

$$f = \dfrac{v}{\lambda} = \left(\sqrt{\dfrac{F}{\mu}}\right)\left(\dfrac{1}{\lambda}\right) = \left(\sqrt{\dfrac{F}{\mu}}\right)\left(\dfrac{1}{2L}\right)$$

Rewriting the above expression:

$f = \dfrac{1}{2L}\sqrt{\dfrac{F}{\mu}}$   $\mu = \dfrac{m}{L}$ is defined as the mass/unit length of the string.

The three laws are summarized by the following statements:
1. The frequency of vibration/oscillation of a stretched string varies *inversely* with the length (*L*) of the string.
2. The frequency varies *directly* with the square root of the stretching force (F) on the string.
3. The frequency varies *inversely* with the square root of the diameter of the string (related to its thickness).

Use the equation to answer two questions:

$f = \dfrac{1}{2L}\sqrt{\dfrac{F}{\mu}}$

In order to double the frequency, and ***keeping the other variables constant in each case***, how does one change
   a) The length of the vibrating string?
   b) The stretching force on the string?

**Appendix C**
Shocking Simulations (Sample questions from the three parts)
Part 1: Build an Atom
   1. Go to http://phet.colorado.edu. Click on HTML5 Sims to get to the list of simulations. From the menu choose the Build an Atom simulation then click on the Atom. You should see the screen at right.
   2. Click on the plus signs to drop down the boxes for *Mass Number, Net Charge,* and *Symbol.*

(To see the SYMBOL you have to select the middle simulation at the bottom of the page)

   3. Drag a proton to the center of the atom (the X). Fill in the following:

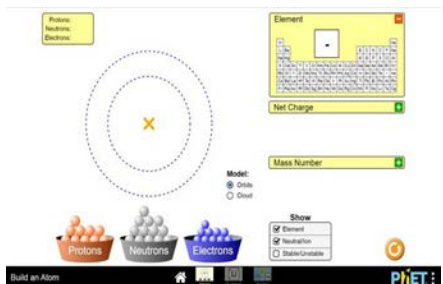

**(When you click the middle simulation at the bottom of the page you must re-drag the proton)

| Symbol | Mass Number | Net Charge |
|--------|-------------|------------|
|        |             |            |

4. What type of atom have you built – a positive ion, a negative ion, or a neutral atom? (1 pt)

5. Add a neutron to the nucleus then fill in the chart below. (3 pts)

| Symbol | Mass Number | Net Charge |
|--------|-------------|------------|
|        |             |            |

6. Does adding a neutron change the net charge of an atom? (1 pt)

7. Add an electron to the orbital ring then fill in the chart below.

| Symbol | Mass Number | Net Charge |
|--------|-------------|------------|
|        |             |            |

8. Does adding an electron change the charge of an atom?

9. What type of atom do you have now - a positive ion, a negative ion, or a neutral atom?

Part Two: Balloons and Static Electricity

17. Go back to the simulations menu for physics and click
18. open the simulation entitled Balloons and Static Electricity.
    Select *Remove Wall.*
    Select *Show no charges*.
    You should see the screen to the right
    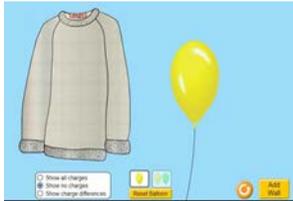
19. Move balloon near the sweatshirt but not touching it and let go. Does the balloon move when you let it go or does it stay put? (1 pt)
20. Does this indicate that the balloon and sweatshirt are charged or neutral?    (1 pt)
21. Now grab the balloon and rub it all over the sweatshirt. Can you tell by looking at the balloon whether it is positive, negative or neutral?    (1 pt)
22. Pull the balloon far away from the sweatshirt and let it go. What does it do?    (1 pt)
23. Does this indicate that the balloon and sweatshirt are charged or neutral?    (1 pt)
24. Do you think the balloon and sweatshirt have the same type of charge or the opposite type of charge?    (1 pt)
25. Click on 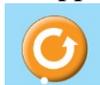 to restart the simulation and *remove the wall* again but this time click on *Show all charges*. Look at the number of positive and negative charges on the balloon and the sweatshirt. Do they look like they are positively charged, negatively charged or
    (approximately) neutral?
    (1 pt)
26. Rub the balloon on the sweatshirt just a little. What happens to the negative charges? (1 pt)
27. Rub the balloon much more on the sweatshirt now. How does this affect the number of charges that are transferred?
    (1 pt)

Part Three: John Travoltage

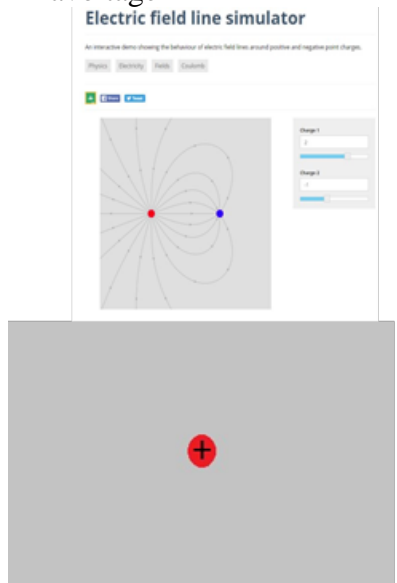

39. Go back to the menu for physics click on the John Travoltage simulation. You should see the screen at right.
40. Be sure his finger is not near the door knob. Rub his foot on the carpet to build up charge on his body. Are these charges positive or negative?   (1 pt)
41. Now bring his finger near the door knob.  Describe what happens to the charges.   (1 pt)
42. Based on your observations, what is happening when you see a spark go between a person's finger and a metal object such as a door knob?   (1 pt)
43. Now, keep his finger near the door knob and rub his foot again on the carpet.  What happens to the charges as you do this – do they build up on his body or do they immediately go into the door knob?
  (1 pt) (2) neatness

## Appendix D
Electric Fields and Forces (Sample questions from both parts)
Part One: Electric Field Lines
1. Using the link in google classroom, go to https://academo.org/demos/electric-field-linesimulator/ . You should see a screen that looks like the one at right.
2. Drag the slider underneath Charge 2 and set it equal to zero. Drag the slider for Charge 1 and set it equal to 1.

3. Sketch this field on the diagram below. Pay careful attention to the direction of
   the arrows.                    (2 pts)

4. Sketch the resulting electric field for these two charges on the diagram below.                        (3 pts)

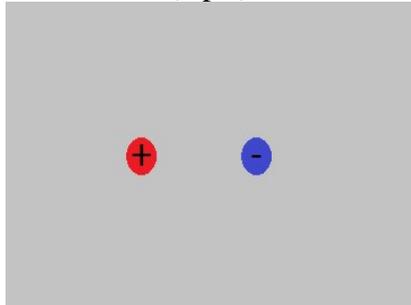

Part Two: Electric Field Hockey

17. Go to http://www.physicsclassroom.com. Click on Physics Interactives on the menu on the left side. Scroll down and select Static Electricity Interactives. On the next screen click

    Put **Launch Interactive** the Charge in the Goal. Select

    Use the Arrows to make the simulation full screen, you should see the screen at right.

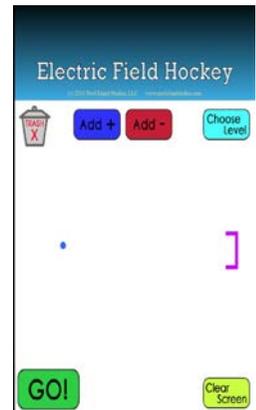

18. The goal of the game is to put the puck into the net using electric fields and forces that you choose. First, we need to do a little warm-up to get you used to the difference between an electric force and an electric field.
19. Click a blue positive charge to add it to the field and drag it to the middle of the playing field. Based on what you learned in Part One of this lab, do the electric field lines match your earlier conclusions? (1 pt)

Electric field lines always point _______ positive charges.

                                                                    (1 pt)

20. Click **GO!**. What happens to the puck? Use details to describe the motion (does it speed up quickly or slowly?)        (2 pt)
21. Does this electric force repel or attract the puck?     (1 pt)
22. What can you conclude about the charge of the puck? (HINT: Do like charges attract or repel?)

European Scientific Journal, ESJ   ISSN: 1857-7881 (Print) e - ISSN 1857-7431
November 2020 edition Vol.16, No.33

(1 pt)

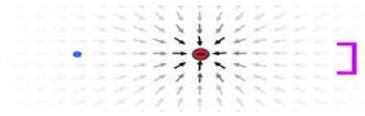

25. Drag the red negative charge closer to the puck and click **GO!**.
Draw a red arrow on the diagram below to depict the direction and the estimated magnitude of the electric force
attracting on the puck from the negative charge.  (2 pt)

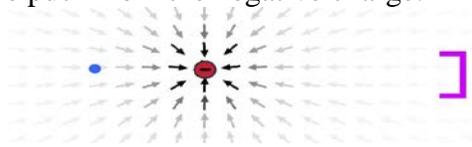

28. How does the size of the force vector compare when the charges are close?

## Appendix E
 Intro to Waves  (Sample questions from the three parts)
 Part One: Transverse Waves
1. Go to http://physics.bu.edu/~duffy/classroom.html.  Click on HTML5 simulations on the menu on the left, scroll down to waves, and select A transverse wave. You should see the screen at right which represents a string with many particles (dots) attached to it.

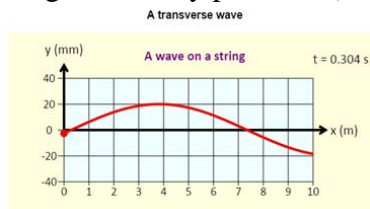

2. Set the *frequency* to 10Hz and the *amplitude* to 40mm.  Describe what you see happening on the screen. [1]
3. Pause the simulation and, on the axes at right, sketch what you see along the string. [1]
4. Change the Amplitude slider bar to its maximum and its minimum values.  What changes about the wave?

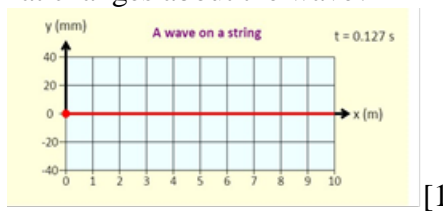

[1

Part Two: Longitudinal Waves
12. On the menu on the left side of the page, select A longitudinal wave. You should see the screen at right. The graphic represents air particles that are being compressed.
13. Watch as the air molecules are alternately compressed and expanded. Which way is this wave of compressions and expansions moving – up and down vertically or left to right horizontally?    [1]

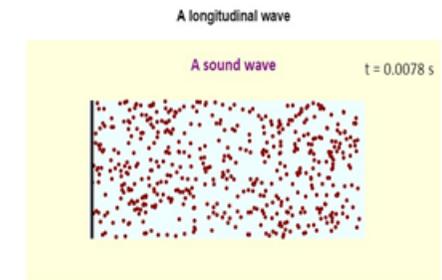

14. Watch a single red molecule as the wave travels. Which way does this single coil/molecule move – up and down vertically or left side to side horizontally?    [1]
15. Compare your answers to the previous two questions. Is the motion of the particle parallel to the motion of the wave?18. Set the *Amplitude* to its highest value and click *PULSE* string. Measure the distance and time and record your data below.

_______________________________________________________________

Part Three: Wave on a String

16. Go to http://phet.colorado.edu. Click on HTML5 to get to the list of simulations. From the menu choose Wave on a String simulation You should see the simulation to the right.

17. Click *No End, Pulse,* and set *Damping* to *None*, *Pulse Width* to .50, and *Tension* to the middle of the slider bar.

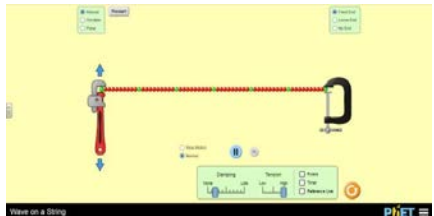

to get the pulse moving

18. Set the Amplitude to its highest value and click PULSE string. Measure the distance and time and record your data below.
Distance: _________________	Time: _________________

19. Calculate the speed of the pulse. Show your calculation in the space below
20. Now move the amplitude bar back and forth from high to low and some in between values and send several pulses down the string. See how long it takes each pulse to get to the end of the string. (HINT: Send a few down right after each other and see who wins the race.) Compare the speed of a high amplitude pulse with the speed of a low amplitude pulse – which is faster or are they the same?

**Appendix F** (Sample questions)
**Optics**
1. The definition of the focal length of a converging lens is the distance to the point where rays initially parallel to the axis meet after passing through the lens. The point is marked by a red circle called the focal point. Why is there a focal point on each side of the lens? Does it make any difference which way light travels through a thin lens?
2. Drag the object back and forth. Describe what you see. What two things are different about the image if the object is closer than the focal length, as compared to when it is further away from the focal length?
3. Use the slider to change the height of the object. How does the height of the image compare to the object height? Does the height of the object change any of your conclusions from the previous question? Explain.
4. For all cases a one ray goes straight through the center of the lens. Why is that? (Hint: Read the introduction.)
5. Carefully describe the other two rays. What happens to a ray that enters the lens parallel to the horizontal axis? What happens to a ray that goes through the focus (if the object is further away from the focus)? What happens to a ray that appears to come from the focus (if the object is closer than the focus)?
6. The previous two questions are about the rules for drawing light rays for a converging lens: 1. Rays parallel to the axis bend and go through the focus on the other side of the lens; 2. Rays going through the focus (or coming from the focus if the object is closer to the focus) bend to exit the lens parallel to the axis; and 3. Rays through the center go straight through without bending. Using these three rules, it is possible to determine where the image will be and how big it will be for any converging lens. Go back and verify these rules. Are they true?
7. Now choose the diverging lens case and experiment. How is it different from the converging case? How does the image size compare with the object size? Is there any case where the image is bigger than the object?

https://www.compadre.org/osp/items/detail.cfm?ID=12399&Attached=1
https://www.compadre.org/osp/EJSS/4476/262.htm?F=1